\documentclass{ws-ijmpa}
\usepackage[super,compress]{cite}
\usepackage{graphicx}
\usepackage{hyperref}
\usepackage{epsfig}
\usepackage{epstopdf}
\usepackage{color}

\begin{document}
\markboth{A.~Yu.~Korchin,  V.~A.~Kovalchuk }{Angular distribution and asymmetries}


\catchline{}{}{}{}{}
%
%%%%%%%%%%%%%%%%%%%%%%%%%%%%%%%%%%%%%%%%%%%%%%%%%%%%%%%%%%%%%%%%%%%%

\title{Angular distribution and asymmetries in the decay  of the polarized charmed baryon $\Lambda_c^+\to K^- \, \Delta^{++} \to K^- \, p \, \pi^+$}

\author{A.~Yu.~Korchin $^{1,2,}$\footnote{korchin@kipt.kharkov.ua}   $^,$\footnote{Corresponding author} ,   
V.~A.~Kovalchuk $^{1,}$\footnote{koval@kipt.kharkov.ua}}

\address{$^1 $NSC ``Kharkiv Institute of Physics and Technology'',  61108 Kharkiv, Ukraine\\
$^2$ V.~N.~Karazin Kharkiv National University,  61022 Kharkiv, Ukraine}

\maketitle

%\begin{history}
 %\received{Day Month Year}
 %\revised{Day Month Year}
 % \end{history}

\begin{abstract}
Angular distribution of the final particles in the decay $\Lambda_c^+\to K^- \, \Delta (1232)^{++} \to K^- \, p \, \pi^+$ 
of the polarized charmed baryon is discussed. Asymmetries are proposed which allow for determination of the 
components of the $\Lambda_c^+$ polarization vector. The precession angle of the polarization in the process 
of baryon channeling in a bent crystal is directly related to these asymmetries. The decay rate and asymmetry parameter for   
the $\Lambda_c^+\to K^- \, \Delta (1232)^{++}$ decay are calculated in the pole model and compared with experiment.

\end{abstract}
 
\ccode{PACS numbers: 13.30.−a, 13.30.Eg, 14.20.−c}

%\tableofcontents
 
%\setcounter{footnote}{0}

\section{\label{sec:Introduction} Introduction}

Measurement of polarization of the charm quark is important for the study of mechanisms
of its production in the QCD processes and for determination of the Lorentz structure
and couplings in particle decays, for example, Higgs boson decay to a $c \bar c$ pair~\cite{Galanti:2015, Korchin:2016}.
The polarization of the $c$ quark determines the polarization of the
$\Lambda^+_c$ baryon \cite{Galanti:2015}, and by investigating the
angular distribution of the decay products of the polarized $\Lambda^+_c$ one can measure
the value of its polarization,  and thus the polarization of the $c$  quark.
Note that the transverse polarization of $\Lambda^+_c$'s from QCD production has already
been seen in the fixed-target experiments NA32~\cite{Jezabek:1992} and E791~\cite{Aitala:2000}.

Another important aspect of the study of $\Lambda^+_c$ polarization is related to
a possibility to measure its magnetic dipole moment (MDM) and electric dipole moment (EDM) using spin precession
in a strong effective magnetic field inside bent crystals~\cite{Baryshevsky:2016, Fomin:2017, Botella:2017, Bagli:2017, Fomin:2019, Mirarchi:2019}.
The motivation here is comparison of experiment with various theoretical calculations of the MDM 
(see, {\it e.g.}, \ \cite{Bernotas:2013, Simonis:2018} and references therein). 
Such measurements may also provide information on the MDM of the charm quark.    

There are various decay modes of the charmed $\Lambda^+_c$ baryon~\cite{PDG:2018}. Our purpose 
is to investigate those modes  for which a measurement of the angular distribution of the final particles is 
more expedient to determine the polarization of the initial $\Lambda^+_c$ baryon. 
Among these modes the hadronic decay $\Lambda^+_c \to K^- \,  p \, \pi^+$, 
which has the largest branching fraction, $6.28 \pm 0.32 \, \%$, is of great interest.   The first
model-independent measurements of the absolute branching fraction
of the $\Lambda^+_c \to K^- \, p \, \pi^+$ decay have been performed by the
Belle~\cite{Belle:2014} and BESIII~\cite{BESIII:2016} collaborations. 
From theoretical point of view the nonleptonic decays of the charmed baryons provide a useful 
environment for studying the interplay between weak and strong interactions.

In the present paper we consider the decay $\Lambda^+_c\to
K^- \, \Delta (1232)^{++}$ which contributes to the decay amplitude 
of $\Lambda^+_c \to K^- \, p \, \pi^+$. Moreover, this process at the quark level 
arises due to mechanism of the $W$-exchange. Therefore, study of
the decay $\Lambda^+_c\to K^- \, \Delta (1232)^{++}$ is important
for the investigation of the $W$-exchange diagrams in the charmed-baryon sector. 

In order to find the matrix element, which determines the decay width 
of $\Lambda^+_c\to K^- \, \Delta (1232)^{++}$ and asymmetry parameter, we apply the  
pole model developed in Refs.~\citen{Xu:1992, Xu_Kamal:1992}. In general, calculation of matrix 
element of the nonleptonic decays of the charmed baryons with $J^P=\frac{1}{2}^+$ involves the factorization 
and non-factorization contributions~\cite{Korner:1992}. The non-factorization contribution can be adequately   
described in the pole model~\cite{Xu:1992, Xu_Kamal:1992}.
Although for description of nonleptonic decay both factorization and non-factorization 
contributions are very often needed, there are cases in which only the pole contribution appears. The process 
$\Lambda_c^+\to K^- \, \Delta^{++}$ belongs to such decays. 

The present paper is organized as follows.   In Sec.~\ref{sec:formalism} definitions and results for 
$\Lambda_c^+\to K^-\, \Delta^{++} \to K^-\, p\, \pi^+$ decay are presented. 
In particular, in Subsec.~\ref{subsec:angle distribution} amplitudes and angular distributions 
are given, and in Subsec.~\ref{subsec:asymmetries} asymmetries are defined 
which determine components of the $\Lambda_c^+$ polarization.  
%%%
In Subsec.~\ref{subsec:precession} the rotation angle of the $\Lambda_c^+$ polarization vector
after baryon passing through a bent crystal is expressed through the asymmetries in the 
$\Lambda_c^+\to K^-\, \Delta^{++} \to K^-\, p\, \pi^+$ decay.   
%%%
In Sec.~\ref{sec:model} the pole model for $\Lambda_c^+\to K^-\, \Delta^{++} $ is described. 
Parameters of the model and theoretical uncertainties are discussed in Subsec.~\ref{subsec:pole model}.
Branching ratio and asymmetry parameter are calculated and compared with experiment. 
As a test of the model, in Subsec.~ \ref{subsec:omega decay} we estimate the rate 
of the weak decay of the strange baryon, $\Omega^- \to K^- \, \Lambda $. 
Concluding remarks are given in Sec.~\ref{sec:conclusions}. 

%%%%%%%%%%%%%%%%%%%%%%%%%%%%%%%%%%%%
%%%%%%%%%%%%%%%%%%%%%%%%%%%%%%%%%%%%

\section{\label{sec:formalism} Amplitudes and angular distributions in  the decay 
$\Lambda_c^+\to K^-\, \Delta^{++} \to K^-\, p \, \pi^+ $ }
\subsection{ \label{subsec:angle distribution} Differential decay rate}

The decay
\begin{equation}\label{eq:0001}
\Lambda_c^+(p)\to K^-(p_2)+\Delta^{++}(p^\prime),
\end{equation}
where $p\,(p^\prime)$ and $p_2$ are the four-momentum of $\Lambda^+_c$
($\Delta (1232)^{++} $) baryon and $K^-$-meson, respectively,
corresponds to the class of transitions $\frac{1}{2}^+\to \frac{3}{2}^++ 0^-$.
Note that throughout this paper integers and fractions with a
superscript $+$ or $-$ will represent $J^P$. The most general form of the transition amplitude is
\begin{equation}\label{eq:0002}
{\cal M} ={\bar
u}_\mu(p^\prime)\left(B-A\gamma_5\right)p_2^\mu\,u(p),
\end{equation}
where $u(p)$ is the Dirac spinor, $u_\mu(p^\prime)$ is the Rarita-Schwinger vector-spinor, 
such that $p^\prime_\mu u^\mu(p^\prime)=0$ and $\gamma_\mu u^\mu(p^\prime)=0$,   
$A$ and $B$ are the Lorentz-invariant amplitudes which have dimension GeV$^{-1}$.
The amplitude $B$ describes the parity-conserving transition ($P$-wave), while $A$ -- parity-violating one ($D$-wave).

In the rest frame of the $\Lambda^+_c$ baryon the helicity amplitudes of the decay $\Lambda_c^+(p,\lambda)\to
K^-(p_2)+\Delta^{++}(p^\prime,\lambda^\prime)$ are defined by the expression
\begin{equation}\label{eq:0003}
F_{0\lambda^\prime\lambda}=-k \, \sqrt{\frac{2\,s}{3\,s^\prime}}\,
\exp{(i\lambda\,\phi_K)}\,d^{\frac{1}{2}}_{\lambda\,-\lambda^\prime}(\theta_K)\,a_{\lambda^\prime},
\end{equation}
where $s=(p_2+p^\prime)^2$ and $s^\prime={p^\prime}^2$. Here
$\lambda$ is the projection of the $\Lambda^+_c$ spin on the axis
$OZ$. The polar angle $\theta_K$ and azimuthal angle $\phi_K$ define the direction of the $K^-$ meson.
Further, $k$ is the momentum of the $K^-$ meson,
$k\equiv|{\vec{p}}_2|=(4s)^{-1/2}\lambda^{1/2}(s,s^\prime,m_2^2)$,
$m_2$ is the mass of $K^-$, $\lambda(a,b,c)$ is the triangle function,
which is symmetrical with respect to all three variables,
$\lambda(a,b,c)=a^2+b^2+c^2-2a\,b-2a\,c-2b\,c$. Finally, $\lambda^\prime$ is the helicity of the $\Delta (1232)^{++} $ isobar.

The helicity amplitudes $a_{\lambda^\prime}$ are related to the invariant amplitudes $A$ and $B$:
\begin{equation}
\label{eq:0004}
a_{\pm 3/2}=0, \qquad a_{1/2}=k_+B+k_-A, \qquad  a_{-1/2}=k_+B-k_-A,
\end{equation}
where
\begin{equation}
\label{eq:0004b}
k_\pm\equiv [ (s^{1/2}\pm {s^\prime}^{1/2})^2-m_2^2]^{1/2}.
\end{equation}

The partial probability $\Gamma(\Lambda_c^+\to
K^-\Delta^{++}) $ for the unpolarized $\Lambda_c^+$ baryon is
\begin{equation}\label{eq:0005}
m_{\Lambda_c}\Gamma(\Lambda_c^+\to K^-\Delta^{++})=\frac{k^3
\sqrt{s}}{12\pi s^\prime}\left(k_+^2|B|^2+k_-^2|A|^2\right),
\end{equation}
where $m_{\Lambda_c}$ is the mass of the $\Lambda_c^+$~\cite{PDG:2018},
while the angular distribution for the polarized $\Lambda_c^+$ baryon reads
\begin{equation}\label{eq:0006}
\frac{1}{\Gamma(\Lambda_c^+\to K^-\Delta^{++})} \frac{d^2\Gamma}{d\cos\theta_K
d\,\phi_K}=\frac{1}{4\pi}\left(1 - \alpha \vec{{\cal P}}\cdot
\hat{\vec{p}}_2\right),
\end{equation}
where $\vec{{\cal P}}$ is the polarization vector of $\Lambda_c^+$, and
$\alpha$ is the asymmetry parameter
\begin{equation}
\label{eq:0007}
\alpha = \frac{|a_{1/2}|^2-|a_{-1/2}|^2}{|a_{1/2}|^2+|a_{-1/2}|^2} 
= \frac{2k_+k_-{\rm Re}(A B^*)}{k_+^2|B|^2+k_-^2|A|^2}.
\end{equation}
Here $\hat{\vec{p}}_2 \equiv \vec{p}_2/|\vec{p}_2|$ is the unit vector chosen along the momentum of
$K^-$ meson.  The sign minus in (\ref{eq:0006}) is related to our choice of the unit vector along the momentum of
the $K^-$ meson. 

In the rest frame of $\Lambda^+_c$, the differential probability of the decay of the polarized $\Lambda^+_c$  
\begin{equation}\label{eq:0008}
\Lambda_c^+(p) \to  K^-(p_2)+\Delta^{++}(p^\prime) 
\to K^-(p_2)+p(p_1) + \pi^+(p_3),
\end{equation}
with $p_1\,(p_3)$ being the four-momentum of the proton ($\pi^+$ meson),
is determined by
\begin{eqnarray}\label{eq:0009}
&&\frac{d^4\,\Gamma}{d\cos\theta_K\,d\phi_K d\cos\theta_p\, d
s^\prime }=\bigl(1 - \alpha  {\cal \vec{P}} \cdot
\hat{\vec{p}}_2\bigr)\left(1+3\cos^2\theta_p\right)
\nonumber \\
&&\times \frac{m_\Delta \,\Gamma_\Delta(s^\prime)}{\left(s^\prime
-m^2_\Delta\right)^2-m^2_\Delta\Gamma^2_\Delta(s^\prime)}\frac{\Gamma(\Lambda^+_c\to
K^-\Delta^{++})}{16\,\pi^2},
\end{eqnarray}
where $m_\Delta$ is the mass of the $\Delta (1232)^{++} $ isobar \cite{PDG:2018}, \ $s^\prime=(p_1+p_3)^2$ is the invariant
mass squared of the $p\,\pi^+$ system. In the rest frame of the $\Delta (1232)^{++}$, $\theta_p$ is the angle between the proton 
momentum and the direction opposite to the momentum of $\Lambda^+_c$ baryon.
$\Gamma_\Delta(s^\prime)$ is the mass-dependent $\Delta (1232)$ width.  
The expression for the latter can be obtained using the decay amplitude of $\Delta \to \pi N$ from Ref.~\citen{Pilkuhn:1979} (Chapter 4). 
Then the width takes the form 
\begin{equation}\label{eq:0010}
\Gamma_\Delta(s^\prime) =\Gamma_\Delta(m^2_\Delta)\left(\frac{k^\prime(s^\prime)}{k^\prime(m^2_\Delta)}\right)^3
\left(\frac{k^\prime_+(s^\prime)}{k^\prime_+(m^2_\Delta)}\right)^2
\frac{m_\Delta}{\sqrt{s^\prime}} \frac{1+{k^\prime}^2(m^2_\Delta)\,r^2_\Delta}{1+{k^\prime}^2(s^\prime)\,r^2_\Delta},
\end{equation}
where $\Gamma_\Delta(m^2_\Delta)$ is the width of the resonance,
$k^\prime(s^\prime)=(4s^\prime)^{-1/2}\lambda^{1/2}(s^\prime,m_1^2,m_3^2)$
is the momentum in the  $p\,\pi^+$ center-of-mass frame,
$k_+^\prime(s^\prime)=(({s^\prime}^{1/2}+ m_1)^2-m_3^2)^{1/2}$,
$m_1\,(m_3)$ is the mass of the proton $p$ ($\pi^+$ meson), and
$k^\prime(m^2_\Delta)$ ($k^\prime_+(m^2_\Delta)$) is
$k^\prime(s^\prime)$ ($k^\prime_+(s^\prime)$) evaluated at the
resonance mass.

The parameter $r_\Delta$ in (\ref{eq:0010}) is the so-called interaction radius, whose value depends on parametrization 
of $\Gamma_\Delta(s^\prime)$.  Sometimes the parametrizations different from (\ref{eq:0010}) are used. 
In Ref.~\citen{Koch:1980} the following form is discussed
\begin{equation}\label{eq:0011}
\Gamma_\Delta(s^\prime)=\Gamma_\Delta(m^2_\Delta)\left(\frac{k^\prime(s^\prime)}{k^\prime(m^2_\Delta)}\right)^3
\frac{1+{k^\prime}^2(m^2_\Delta)\,r^2_\Delta}{1+{k^\prime}^2(s^\prime)\,r^2_\Delta},
\end{equation}
then $r_\Delta=1.11\pm0.02$ fm. Another parametrization of $\Gamma_\Delta(s^\prime)$ was suggested 
in \cite{Pilkuhn:1979}:
\begin{equation}\label{eq:0012}
\Gamma_\Delta(s^\prime)=\Gamma_\Delta(m^2_\Delta)\left(\frac{k^\prime(s^\prime)}{k^\prime(m^2_\Delta)}\right)^3
\frac{1+{k^\prime}^2(m^2_\Delta)\,r^2_\Delta}{1+{k^\prime}^2(s^\prime)\,r^2_\Delta}\,\frac{m_\Delta}{\sqrt{s^\prime}}
\end{equation}
with $r_\Delta= 1.03\pm0.02$ fm. Ref.~\citen{Pilkuhn:1979} also gives a 
few reasons for neglecting the multiplier ${k^\prime_+ (s^\prime) }^2 / {k^\prime_+(m_\Delta^2)}^2 $ 
in Eq.~(\ref{eq:0012}), while at the same time it is emphasized that these reasons are not quite convincing. 

In the analyses of the pion-nucleon scattering all these parameterizations lead to practically the same mass and width of the $\Delta$-resonance, 
and affect only the shape of the resonance curve. 
Concerning experimental studies of the decay $\Lambda_c^+ \to K^- \, \Delta^{++} \to K^- \, p\, \pi^+$, it would be of interest   
to investigate sensitivity of asymmetry parameter to the parameterization of $\Gamma_\Delta(s^\prime)$.        

The fully differential angular distribution for the 3-body decay (\ref{eq:0008}) is given by
\begin{eqnarray}\label{eq:0013}
 W(\theta_K, \phi_K,\theta_p)
& \equiv&
\frac{d^4\,\Gamma}{d\cos\theta_K\,d\phi_K\,
d\cos\theta_p\,d s^\prime} \, \Big/ \, \frac{d \Gamma}{d s^\prime} \nonumber \\
& = & \frac{1}{16\,\pi}\bigl(1 - \alpha \vec{{\cal P}}\cdot
\hat{\vec{p}}_2\bigr)\left(1+3\cos^2\theta_p\right),
\end{eqnarray}
where the distribution over the $\Delta$  invariant mass is
\begin{equation}\label{eq:0014}
\frac{d\,\Gamma}{d s^\prime }=\frac{1}{\pi}\frac{m_\Delta
\,\Gamma_\Delta(s^\prime)}{\left(s^\prime
-m^2_\Delta\right)^2-m^2_\Delta\Gamma^2_\Delta(s^\prime)}\,\Gamma(\Lambda^+_c\to
K^-\Delta^{++}).
\end{equation}

%%%%%%%%%%%%%%%%%%%%%%%%%%%%%%%%%%%
%%%%%%%%%%%%%%%%%%%%%%%%%%%%%%%%%%%

\subsection{ \label{subsec:asymmetries}   One-dimensional angular distributions and asymmetries}

The one-dimensional angular distributions in $\cos\theta_K$ and
$\phi_K$ are simply
\begin{equation}
W_{\theta_K}(\cos\theta_K)  \equiv  \frac{d^2\,\Gamma}{d\cos\theta_K d s^\prime} \, \Big/ \, \frac{d\,\Gamma}{d s^\prime}  
= \frac{1}{2} \bigl(1 - \alpha {\cal P}_z\cos\theta_K\bigr) 
\label{eq:0015}
\end{equation}
and
\begin{equation}
W_{\phi_K}(\phi_K) \equiv  \frac{d^2\,\Gamma}{d\phi_K d s^\prime}\, \Big/ \, \frac{d\,\Gamma}{d s^\prime} 
=\frac{1}{2\pi} - \frac{\alpha}{8} \bigl({\cal P}_x\cos\phi_K+{\cal P}_y\sin\phi_K\bigr). 
\label{eq:0016}
\end{equation}

Study of the distribution (\ref{eq:0015}) allows one to measure the product $\alpha {\cal P}_z$. Indeed, we
can define the forward-backward asymmetry of the $K^-$ mesons
\begin{equation}\label{eq:0017}
A_{\rm FB}=\frac{F-B}{F+B},
\end{equation}
where
\begin{equation}
F\equiv \int_0^1 W_{\theta_K} (\cos\theta_K) \, d\cos\theta_K,  \qquad 
B\equiv \int_{-1}^0 W_{\theta_K} (\cos\theta_K) \, d\cos\theta_K. 
\label{eq:0018}
\end{equation}
This asymmetry is equal to
\begin{equation}\label{eq:0019}
A_{\rm FB}= - \frac{\alpha}{2}{\cal P}_z,
\end{equation}
and its measurement allows one to find the $z$ component ${\cal P}_z$ of the polarization vector once the value of $\alpha$ is known.

Measurement of the angular distribution in the azimuthal angle
$\phi_K$ (\ref{eq:0016}) allows one to determine the components ${\cal P}_x$ and ${\cal P}_y$:
\begin{equation}
A_x \equiv \Bigl(\int\limits_{0}^{\pi/2}d\,\phi_K-\int\limits_{\pi/2}^{3\pi/2}d\,\phi_K+\int\limits_{3\pi/2}^{2\pi}d\,\phi_K
 \Bigr)W_{\phi_K} (\phi_K) 
= - \frac{\alpha}{2}{{\cal P}_x}, \label{eq:0020}
\end{equation}
\begin{eqnarray}
A_y \equiv\Bigl(\int\limits_{0}^{\pi}d\,\phi_K-\int\limits_{\pi}^{2\pi} d\,\phi_K
 \Bigr) W_{\phi_K} (\phi_K) = - \frac{\alpha}{2}{{\cal P}_y} . \label{eq:0021}
\end{eqnarray}

%%%%%%%%%%%%%%%%%%%%%%%%%%%%%%%%%%%
%%%%%%%%%%%%%%%%%%%%%%%%%%%%%%%%%%%

\subsection{ \label{subsec:precession} Application to precession of the $\Lambda_c^+$ polarization in bent crystals }

In general case for measurement of the polarization components one has to know the asymmetry
parameter $\alpha$ and magnitude of polarization ${\cal P}$.
However, for measurement of the magnetic dipole moment (MDM) and electric dipole moment (EDM) of a short-lived fermion
using technique of the bent crystals (see Refs.~\citen{Fomin:2017, Botella:2017, Fomin:2018}) it is sufficient to determine only 
the rotation angles of the polarization vector.
We will show that in this case one can directly use the asymmetries introduced in
Subsec.~\ref{subsec:asymmetries} without knowledge of parameter $\alpha$ and magnitude of polarization.

Let us assume that in front of the bent crystal the initial polarization vector is oriented along the OX axis,
$\Lambda_c^+$ baryon moves along the OZ axis (see Fig.~1 in Ref.~\citen{Fomin:2017}), and the average electric field
in the crystal is directed along the OX axis. Thus  the three-vectors of initial polarization, baryon velocity and
electric field have the components
\begin{equation}
\vec{{\cal P}}_{in} = ({\cal P},\,  0,\,  0), \qquad  \vec{v} = (0,\,  0,\,  v), \qquad \vec{E} = (-E,\,  0,\, 0).
\label{eq:00211}
\end{equation}

In general the magnetic dipole moment (MDM) and electric dipole moment (EDM) of the baryon are written as
\begin{equation}
\vec{\mu} = g\, \frac{q }{2 m c} \, \vec{S}, \qquad  \quad \vec{d} = \eta \,  \frac{q }{2 m c} \, \vec{S},
\label{eq:00212}
\end{equation}
where $q$ is the electric charge of the baryon with the mass $m$ and 
spin $\vec{S} = \tfrac{\hbar}{2} \vec{\sigma}$, \ $g$ is the gyromagnetic factor ($g$-factor) and
$\eta$ is a similar factor for the EDM.

After passing the crystal, the spin, or the polarization vector $\vec{{\cal P}} = \tfrac{2}{\hbar}  \langle  \vec{S} \rangle$,
rotates around the axis which is determined by the equations for the spin precession in external electric and magnetic fields
\cite{Berestetskii:1982, Bargmann:1959, Jackson:1999,Lyuboshits:1980,Kim:1983}. 
In particular, for the electric field
in (\ref{eq:00211}), which is orthogonal to the velocity at any moment of time,  $\vec{E}  \vec{v}=0$, one finds the angular
velocity of the polarization rotation
\begin{eqnarray}
&& \vec{\Omega} = ( {\omega}^\prime, \,  -{\omega}, \, 0), \label{eq:00213}  \\
&& \omega =  \gamma  \omega_v \ (a -\frac{g}{2 \gamma^2} + \frac{1}{\gamma}), \qquad \;
\omega^\prime =  \gamma  \omega_v  \ \frac{\eta \, v}{2 c}. \nonumber
\end{eqnarray}
Here $a = \tfrac{1}{2}(g -2)$ is the anomalous magnetic moment of the baryon,
$\gamma = (1 - v^2/c^2)^{-1/2}$ is the Lorentz factor,
and the angular velocity of the momentum rotation $\vec{\omega}_v$ is defined through
\begin{equation}
\vec{\omega}_v = (0, \, - \omega_v, \, 0),
\qquad  \quad  \omega_v = \frac{q E}{m \gamma v} = \frac{v}{R},
 \label{eq:00214}
\end{equation}
where $R$ is curvature of the crystal.

Integration of Eq.~(\ref{eq:00213}) over time assuming constant velocity leads to relations 
\begin{eqnarray}
&& \vec{\Phi} =  ({\theta}^\prime, \,  - {\theta}, \, 0),   \label{eq:00215}  \\
&& \theta =  \gamma  \theta_v \, (a -\frac{g}{2 \gamma^2} + \frac{1}{\gamma}) ,
\qquad  \quad \theta^\prime =  \gamma  \theta_v  \,\frac{\eta \, v}{2 c}, \nonumber \\
&& \vec{\theta}_v = (0, \, - \theta_v, \, 0),   \qquad  \quad \theta_v = \frac{L}{R},
 \nonumber
\end{eqnarray}
where $L$ is the arc length that baryon passes in the channeling regime.  The crystal length and crystal curvature 
for the $\Lambda_c^+$ baryon have been analyzed and optimized in Refs.~\citen{Fomin:2017, Fomin:2019}. 

Eqs.~(\ref{eq:00215}) imply that the polarization vector rotates around the unit vector
$\vec{n}$ \ by the angle $\Phi$ which are defined as follows
\begin{equation}
\vec{n} = (\frac{\theta^\prime}{\Phi}, \, - \frac{\theta}{\Phi},\, 0),  \qquad \quad  \Phi=\sqrt{\theta^2 + {\theta^{\prime}}^2 }.
 \label{eq:00216}
\end{equation}

Then the components of the baryon polarization vector after passing the crystal are 
\begin{eqnarray}
&& \vec{{\cal P}}_{fin} = ({\cal P}_x, \, {\cal P}_y, \, {\cal P}_z), \label{eq:00217} \\
&& {\cal P}_x =  {\cal P} \,  \frac{1}{\Phi^2} ( \theta^2  \cos \Phi  + {\theta^\prime}^2 ), \nonumber \\
 && {\cal P}_y =  {\cal P} \, \frac{\theta {\theta^\prime}}{\Phi^2} \, (\cos \Phi -1) , \nonumber \\
&& {\cal P}_z =  {\cal P} \, \frac{\theta}{\Phi} \, \sin \Phi. \nonumber
\end{eqnarray}

The angles $\theta$  and $\theta^\prime$ are determined from ratios of the asymmetries
(\ref{eq:0019})-(\ref{eq:0021}):
\begin{equation}
\frac{A_x}{A_{\rm FB}} = \frac{ \theta^2 \cos \Phi  + {\theta^\prime}^2 }{ \theta \, \Phi \sin \Phi}, \qquad \qquad 
 \frac{A_y}{A_{\rm FB}} = \frac{\theta^\prime (\cos \Phi -1)}{ \Phi \sin \Phi }.
\label{eq:00218}
\end{equation}
It is seen that the parameter $\alpha$ and the magnitude of the polarization ${\cal P}$ do not enter
these equations. The angles $\theta$ and $\theta^\prime$, and correspondingly the anomalous magnetic moment $a$ 
and electric dipole moment $\eta$ can be directly found from ratios of the asymmetries in Eqs.~(\ref{eq:00218}). 

Under the assumption that the angle $\theta^\prime$ is small compared to the
angle $\theta$, one has $\Phi \approx \theta$ and Eqs.~(\ref{eq:00218}) simplify:
 \begin{equation}
\frac{A_x}{A_{\rm FB}} =  \cot \theta,  \qquad \quad 
 \frac{A_y}{A_{\rm FB}} = \frac{\theta^\prime \, (\cos \theta - 1) }{\theta \,  \sin \theta}
\label{eq:00219}
\end{equation}
and it is seen, in particular, that the asymmetry $A_y$ is not zero only if baryon has a nonzero EDM.

Eqs.~(\ref{eq:00218}) and (\ref{eq:00219}) may be useful in measurements  
of MDM and EDM of the short-lived baryons using bent crystals at 
CERN~\cite{Baryshevsky:2016,Fomin:2017,Botella:2017,Bagli:2017,Fomin:2019,Mirarchi:2019}.

%%%%%%%%%%%%%%%%%%%%%%%%%%%%%%%%%%%%%%%%%%%%%%%%%%%%%%%%%%%%%%%%%%%%%
%%%%%%%%%%%%%%%%%%%%%%%%%%%%%%%%%%%%%%%%%%%%%%%%%%%%%%%%%%%%%%%%%%%%%
%%%%%%%%%%%%%%%%%%%%%%%%%%%%%%%%%%%%%%%%%%%%%%%%%%%%%%%%%%%%%%%%%%%%%

\section{\label{sec:model} Model calculation of decay rate and asymmetry}

\subsection{\label{subsec:pole model} Pole model for the decay
$\Lambda_c^+\to K^-\, \Delta (1232)^{++} $}

\begin{figure}[tbh]
\begin{center}
\includegraphics[width=0.95\textwidth]{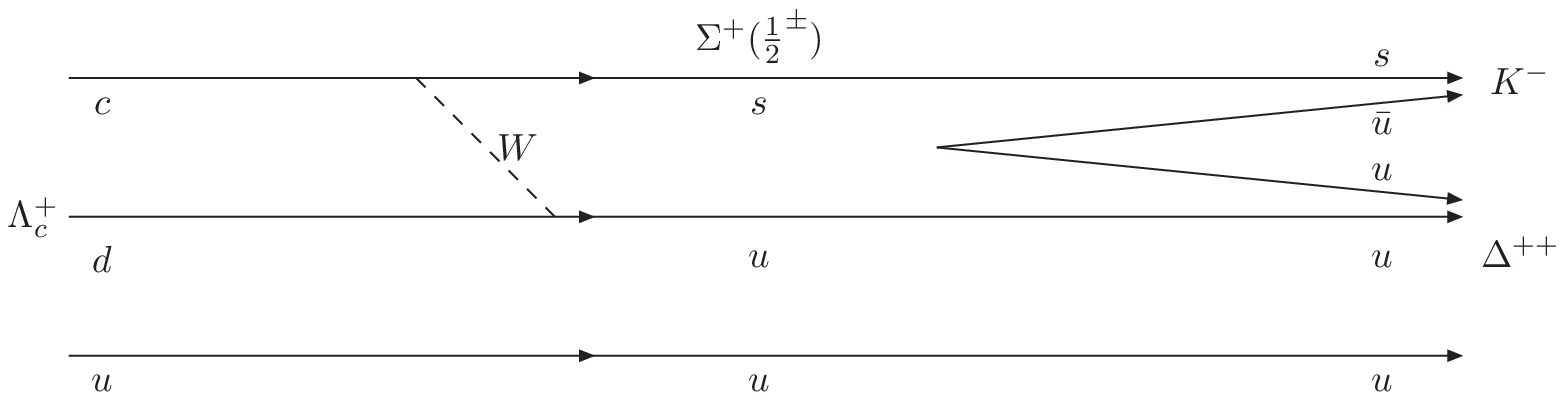}
\end{center}
\caption{Diagram of the process $\Lambda_c^+ \to K^-\, \Delta^{++}$ at the  
quark level. }
\label{fig:Diagram1}
\end{figure}

\begin{figure}[tbh]
\begin{center}
\includegraphics[width=0.60\textwidth]{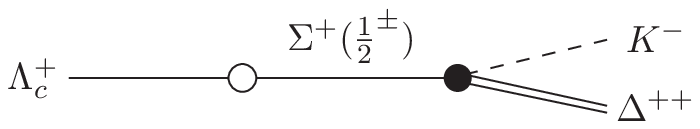}
\end{center}
\caption{$s$-pole amplitude of the process $\Lambda_c^+\to  \Sigma^+ (\tfrac{1}{2}^\pm) \to K^-\, \Delta^{++}$. }
\label{fig:Diagram2}
\end{figure}

In the pole model \cite{Xu:1992, Xu_Kamal:1992} the charmed baryon $\Lambda_c^+ (udc)$ due to the weak interaction mediated by 
the $W$-boson exchange transforms into the intermediate baryons $\Sigma^+ (uus)$ of positive or 
negative parity,  $\Sigma^+(J^P=\tfrac{1}{2}^+)$ or $\Sigma^+(J^P=\tfrac{1}{2}^-)$. Further, the strong 
interaction induces the decay of the $\Sigma^+  (uus)$ to the  
state $ K^- (s \bar{u}) \,  + \, \Delta^{++} (uuu)$ (see Fig.~\ref{fig:Diagram1}). 
This two-step mechanism at the hadronic level can be described by the $s$-pole 
amplitude in Fig.~\ref{fig:Diagram2}.

%%%%%%%%%%%%%%%%%%%%%%%
%%%%%%%%%%%%%%%%%%%%%%%

%%%%%%%%%%%%%%%%%%%%%%%
%%%%%%%%%%%%%%%%%%%%%%%

To describe the transition $\Sigma^+ \to K^- \, \Delta^{++}$ one can use the following interaction Lagrangians
\begin{equation}
{\cal L}_{\Sigma ({1}/{2}^\pm) K  \Delta} = g_{\pm} \,
\bigl( \bar{\Delta}^\mu \, \vartheta_{\pm} \, \Sigma ({1}/{2}^\pm) \bigr)   \partial_\mu K +  \mathrm{ H.c.}
\label{eq:0022} \\
\end{equation}
with $\vartheta_+ =1$ and $\vartheta_- =\gamma_5$.

For the matrix element of the weak effective Hamiltonian ${\cal H}_W$ between the states of $\Lambda_c^+$ and $\Sigma^+ ({1}/{2}^\pm)$ 
one can write the following~\cite{Xu_Kamal:1992}
\begin{eqnarray}
&& \langle \, \Sigma^+ ({1}/{2}^+) | {\cal H}_W^{pc} | \Lambda_c^+ \, \rangle = h_+ \, \bar{u}_{\Sigma^+ ({1}/{2}^+)} \, u_{\Lambda_c^+} ,
\label{eq:0023} \\
&& \langle \, \Sigma^+ ({1}/{2}^-) | {\cal H}_W^{pv} | \Lambda_c^+ \, \rangle = h_- \, \bar{u}_{\Sigma^+ ({1}/{2}^-)} \, u_{\Lambda_c^+}.
\label{eq:0024}
\end{eqnarray}
In Eqs.~(\ref{eq:0022})-(\ref{eq:0024}) we used the shorthanded notation
\begin{equation}
g_+ \equiv g_{\Delta^{++} K^\pm \Sigma^+ ({1}/{2}^+) },  \quad  g_- \equiv g_{\Delta^{++} K^\pm \Sigma^+ ({1}/{2}^-) },  
\quad   h_\pm \equiv h_{\Sigma^+ ({1}/{2}^\pm) \Lambda_c^+}.
\label{eq:0025}
\end{equation}

Using the above definitions we find the amplitudes $B$ and $A$ in Eq.~(\ref{eq:0002}):
\begin{eqnarray}
&& B= \sum_{j=1} \, \frac{h_{+, j} \, g_{+, j}}{ \sqrt{s}  - M_{\Sigma^{+}_j ({1}/{2}^+)} + \tfrac{i}{2} \Gamma_{\Sigma^{+}_j ({1}/{2}^+)} }, 
\label{eq:0026} \\
&& A  = \sum_{j=1} \, \frac{h_{-, j} \, g_{-, j}}{\sqrt{s} - M_{\Sigma^{+}_j ({1}/{2}^-)} + \tfrac{i}{2} \Gamma_{\Sigma^{+}_j ({1}/{2}^-)} }
\label{eq:0027}
\end{eqnarray}
in terms of the masses and the total decay widths of the intermediate baryons $\Sigma^{+}_j ({1}/{2}^\pm)$. 
The sums run over the contributing states. 

Let us discuss values of the model parameters. These parameters were suggested in Ref.~\citen{Xu_Kamal:1992}; 
here we update these values using the present experimental information~\cite{PDG:2018}.

For the contribution from the lowest-mass positive-parity $\Sigma^+$ baryon with the mass $1189.37 \pm 0.07$ MeV we need the constants 
$g_{\Delta^{++}  K^\pm \Sigma^+ }$ and $h_{\Sigma^+  \Lambda_c^+}$. 
To find the former, one can apply the $SU(3)$ symmetry relations for the product of constants $g_{B^\prime P B} \, f_P$, 
where $f_P$ is the constant of the weak decay of the pseudoscalar meson $P$, baryon $B$ belongs to the $SU(3)$ 
octet $\tfrac{1}{2}^+$ and $B^\prime$ -- to the $SU(3)$ decuplet $\tfrac{3}{2}^+$. The generalized Goldberger-Treiman 
relation for the axial-vector current form factor and the $SU(3)$ symmetry  gives~\cite{Xu_Kamal:1992}:
\begin{eqnarray}
&& g_{\Delta^{++}  K^+ \Sigma^+ } f_K= g_{\Omega^{-}  K^- \Xi^0} f_K \nonumber \\
&& = \sqrt{3} \, g_{\Xi (1530)^{-} \pi^- \Xi^0} f_\pi = \sqrt{3} \, g_{\Xi (1530)^{0} K^- \Sigma^+} f_K =  \ldots
\label{eq:0028}
\end{eqnarray}

Then one can find the constant $g_{\Delta^{++}  K^+ \Sigma^+}$ for the kinematically forbidden 
decay $\Delta^{++} \to  K^+ \Sigma^+ $ through the constant $g_{\Xi (1530)^{-} \pi^- \Xi^0}$.  
The experiment~\cite{PDG:2018} gives the decay width 
$\Gamma (\Xi (1530)^{-} \to \pi \Xi)=9.9^{+1.7}_{-1.9}$ MeV.  
Taking into account the error on the width, and using the ratio~\cite{Rosner:2010}  ${f_K}/{f_\pi} = 1.197 \pm 0.002 \pm 0.006 \pm 0.001$ 
(central value), we get 
\begin{equation}
g_{\Delta^{++}  K^+ \Sigma^+ }=     8.78 \pm 0.80 \; {\rm GeV}^{-1}.
\label{eq:0029}
\end{equation}
The constant $h_{\Sigma^+  \Lambda_c^+} = 0.8 \times 10^{-7}$ GeV 
is taken from~\cite{Xu_Kamal:1992}.

There exist a few negative-parity baryons $\Sigma^+ (J^P= \tfrac{1}{2}^-)$, in particular, $\Sigma (1620)^+$ and $\Sigma (1750)^+$, 
which can contribute to the amplitude $A$.  
The constant $ g_{\Delta^{++}  K^\pm \Sigma (1620)^+} $ can be obtained assuming $SU(3)$ relations like   
Eqs.~(\ref{eq:0028})
\begin{equation}
g_{\Delta^{++}  K^+ \Sigma (1620)^+ } f_K= g_{\Delta^{++} \pi^+ N (1535)^+ } f_\pi  
= \sqrt{3} \, g_{\Xi (1530)^0 K^- \Sigma (1620)^+  } f_K =  \ldots
\label{eq:0030}
\end{equation}
for the baryons of decuplet $\tfrac{3}{2}^+$ and octet $\tfrac{1}{2}^-$. Similar relations hold for the next 
octet $\tfrac{1}{2}^-$ containing $\Sigma (1750)^+ $ and $N (1650)^+$.

Experimental information on the rates of $N (1535)^+ \to \pi \Delta$ and $N (1650)^+ \to \pi \Delta$ decays exists~\cite{PDG:2018}, 
however, the branching fractions are not precise: \
${\rm B}(N (1535)^+ \to \pi \Delta) = 1-4 \, \%$ and ${\rm B}(N (1650)^+ \to \pi \Delta) = 6-18 \, \%$.  
We can also use experimental results~\cite{Sokhoyan:2015}    
${\rm B}(N (1535)^+ \to \pi \Delta) = 2.5 \pm 1.5 \, \%$ and ${\rm B}(N (1650)^+ \to \pi \Delta) = 12 \pm 6 \, \%$, 
and the total decay widths~\cite{PDG:2018} $\Gamma_{tot}(N (1535)^+)=150$ MeV and  $\Gamma_{tot}(N (1650)^+)=125$ MeV. 
Then the needed couplings with uncertainties are   
\begin{eqnarray}
&& g_{\Delta^{++}  K^+ \Sigma (1620)^+ } =   8.98 \pm 2.69 \; {\rm GeV}^{-1},
\label{eq:0031} \\
&& g_{\Delta^{++}  K^+ \Sigma (1750)^+ } = 6.89 \pm 1.72 \; {\rm GeV}^{-1}.
\label{eq:0032}
\end{eqnarray}
The values of the constants $h_{\Sigma (1620)^+ \Lambda_c^+} = 0.32 \times 10^{-7}$ GeV and 
$h_{\Sigma (1750)^+ \Lambda_c^+} = 0.79 \times 10^{-7}$ GeV are taken from Ref.~\citen{Xu_Kamal:1992}.

%%%%%%%%%%%%%%%%%%% TABLE %%%%%%%%%%%%%%%%%%%%%
%%%%%%%%%%%%%%%%%%%%%%%%%%%%%%%%%%%%%%%%%%%%
\begin{table}[htb]
\tbl{ The partial decay width $\Gamma_0 \equiv \Gamma(\Lambda_c^+ \to K^- \Delta^{++})$ 
(in units $10^{10} \ s^{-1}$) and asymmetry parameter $\alpha$.
The first column shows two variants of calculation with $A=A_1+A_2$ and $A=A_1 -A_2$, the second column -- our calculation,
the third column -- calculation of Ref.~\citen{Xu_Kamal:1992}, 
the fourth column -- prediction of Refs.~\citen{Korner:1992, Konig:1993}, and 
the last column shows data for the decay width~\cite{PDG:2018} and 
the asymmetry parameter~\cite{Aitala:2000,Botella:2017}. }
{\label{tab:width_asymmetry}
%\begin{center}
\begin{tabular}{|c |c | c | c | c|}
  \hline
 $A$  &  $\Gamma_0$  &  $\Gamma_0 $ \ (Ref.~\citen{Xu_Kamal:1992}) &   $\Gamma_0 $  \ (Ref.~\citen{Korner:1992})   &     $\Gamma_{0, \, {exp}}$ \ (Ref.~\citen{PDG:2018})    \\
  &  $\alpha$  & $\alpha$ \ (Ref.~\citen{Xu_Kamal:1992}) &    $\alpha$ \ (Refs.~\citen{Korner:1992, Konig:1993})   &    $\alpha_{exp}$ \  (Refs.~\citen{Aitala:2000,Botella:2017})       \\
   \hline
 $A_1 + A_2 $ &  $ 6.38 \pm 1.08 $   & 5.2 &    &    \\
         &  $ 0.87 \pm 0.09 $   &  0.43 &  &    \\
         &                       &   &       13.5                      &  $ 5.4 \pm 1.26 $     \\
			  &                       &   &  0.0    &  $-0.67 \pm 0.30 $    \\					
 $A_1 - A_2$   & $ 5.03 \pm 0.90 $ & 5.0 &   &             \\
               & $ -0.48 \pm 0.20 $  & 0.0 &    &  \\
\hline
\end{tabular}}
%\end{center}
\end{table}
%%%%%%%%%%%%%%%%%%%%%%%%%%%%%%%%%%%%%%%

Results of our calculation of the $\Lambda_c^+ \to K^- \Delta^{++}$ decay rate (\ref{eq:0005}) and 
asymmetry parameter (\ref{eq:0007}) are presented in Table~\ref{tab:width_asymmetry}. 
In calculation we set $\sqrt{s}=m_{\Lambda_c}$ and $\sqrt{s^\prime}=m_\Delta$. The estimated uncertainties on 
the calculated width and asymmetry parameter are induced by the uncertainties on the coupling constants 
$g_{\Delta^{++}  K^+ \Sigma^+ }, \; g_{\Delta^{++}  K^+ \Sigma (1620)^+ }$ 
and $g_{\Delta^{++}  K^+ \Sigma (1750)^+ }$.  
We do not include uncertainties related to the couplings $h_\pm$, particle masses, widths of $\Sigma (1620)^+$ and $\Sigma (1750)^+$,    
and possible inaccuracy of the $SU(3)$ relations (\ref{eq:0028}) and (\ref{eq:0030}).
  
The largest uncertainty of the calculation comes from the relative sign of amplitudes $A_1$ and $A_2$ in Eq.~(\ref{eq:0027}), 
corresponding to the $\Lambda_c^+ \to \Sigma  (1620)^+ \to K^- \Delta^{++} $  and 
$\Lambda_c^+ \to \Sigma (1750)^+ \to K^- \Delta^{++} $ transition, respectively.  
A negative sign between these amplitudes gives rise to decay width and asymmetry 
parameter, which are in reasonable agreement with experiment ({\it cf.} the second and fifth columns in Table~\ref{tab:width_asymmetry}). 

In Table~\ref{tab:width_asymmetry} we also show results of calculation~\cite{Xu_Kamal:1992} and prediction of the spectator 
quark model~\cite{Korner:1992}.   In the latter approach the decay $\Lambda_c^+ \to K^- \Delta^{++}$ is a parity-conserving transition
and the asymmetry parameter vanishes~\cite{Korner:1992, Konig:1993}.  Note that the experimental value of the asymmetry parameter 
in Table~\ref{tab:width_asymmetry} was calculated in Ref.~\citen{Botella:2017}, where the measured amplitudes from Ref.~\citen{Aitala:2000} 
were used.

%%%%%%%%%%%%%%%%%%%%%%%%%%%%%%%
%%%%%%%%%%%%%%%%%%%%%%%%%%%%%%%

\subsection{\label{subsec:omega decay} Test of the model for decay $\Omega^- \to K^-\, \Lambda $}

Following Ref.~\citen{Xu_Kamal:1992}, we test the pole model for the weak decay of the strange baryon,
$\Omega^- \to K^-\, \Lambda$.  The latter decay has some similarities with the $\Lambda_c^+\to K^- \, \Delta^{++}$ decay. 
Indeed, in framework of the pole model, \ $\Omega^- \to K^-\, \Lambda$ proceeds via the strong-interaction 
process \ $\Omega^- (sss) \to K^- (s \bar{u}) \, \Xi^0 (uss)$ followed by the conversion \ $\Xi^0 (uss) \to \Lambda (dus)$ 
due to the $W$-exchange. This mechanism corresponds to the $u$-pole amplitude in terminology of the pole model. 

The decay $\Omega^- \to K^-\, \Lambda $ is the transition $\tfrac{3}{2}^+ \to \tfrac{1}{2}^+ +0^- $, and
the decay width reads in terms of the amplitudes $B$ and $A$ 
\begin{equation}
\Gamma(\Omega^- \to K^- \Lambda) = \frac{k^3}{24 \pi m_{\Omega}^2 } \left(k_+^2|B|^2+k_-^2|A|^2\right),  
\label{eq:0034}
\end{equation}
where $k$ is the momentum of $\Lambda$ (or $K^-$) in the $\Omega^-$ rest frame and 
$k_\pm \equiv  ( (m_\Omega  \pm m_\Lambda)^2 - m_K^2 )^{1/2}$.  

The parity-conserving amplitude $B$ has the form
\begin{equation}
B= \frac{h_{\Lambda \, \Xi^0} \, g_{\Omega^- K^- \Xi^0}}{ M_\Lambda  - M_{\Xi^{0} } },
\label{eq:0035}
\end{equation}
where we keep contribution from the lowest-mass baryon $\Xi^0$ with $J^P=\tfrac{1}{2}^{+}$ and mass $1314.86 \pm 0.20$ MeV.  
The coupling $g_{\Omega^- K^- \Xi^0}$ is equal to $g_{\Delta^{++}  K^+ \Sigma^+ }$ due to (\ref{eq:0028}), 
and is given in (\ref{eq:0029}). The value of the constant $h_{\Lambda \, \Xi^0}$ is calculated from Eqs.~(22)
in Ref.~\citen{Stech:1990}. It turns out to be $h_{\Lambda \, \Xi^0} = -0.88 \times 10^{-7}$ GeV.

The parity-violating amplitude $A$ can be written similarly to Eq.~(\ref{eq:0035}) with the change 
$\Xi^0 \to \Xi^0 ({1}/{2}^-)$.  
However, the magnitude of $A$ is difficult to estimate as the negative-parity baryons  $\Xi^0 $ with $J^P=\tfrac{1}{2}^{-}$ 
are not listed in the latest Review of Particle Physics~\cite{PDG:2018}.
In any case, the contribution from the amplitude $A$ to the decay width of $\Omega^- \to K^-\, \Lambda $ is suppressed because 
of the smallness of the factor $k_{-}$ in (\ref{eq:0034}). Indeed, $ k_-^2 / k_+^2 \approx 0.009$. 
One can also expect that the asymmetry parameter (\ref{eq:0007}) is small. 

%The coupling $g_{\Omega^- K^- \Xi^0}$ is equal to $g_{\Delta^{++}  K^- \Sigma^+ }$ due to (\ref{eq:0028}), 
%and is given in (\ref{eq:0029}). The value of the constant $h_{\Lambda \, \Xi^0}$ is calculated from Eqs.~(22)
%in Ref.~\citen{Stech:1990}. It turns out to be $h_{\Lambda \, \Xi^0} = -0.88 \times 10^{-7}$ GeV.

The decay width calculated using Eqs.~(\ref{eq:0034}) and (\ref{eq:0035}) is \  
$\Gamma (\Omega^- \to K^-\, \Lambda) = (7.64 \pm 1.39) \times 10^{9}$ $s^{-1}$,  where  
the estimated uncertainty comes from uncertainty on the coupling $g_{\Omega^- K^- \Xi^0}$. 
This result can be compared with the data~\cite{PDG:2018}: \ $\Gamma (\Omega^- \to K^-\, \Lambda) _{exp}= (8.26 \pm 0.14) \times 10^{9}$ $s^{-1}$. 
The experimental value of asymmetry parameter is very small, $\alpha_{exp}=0.0154 \pm 0.0020$~\cite{PDG:2018}.      
It is seen that the calculation in this simple model does not contradict the experiment. This gives confidence in the predictive power 
of the pole model.

%%%%%%%%%%%%%%%%%%%%%%%%%%%%%%%%
%%%%%%%%%%%%%%%%%%%%%%%%%%%%%%

\section{ \label{sec:conclusions} Conclusions}

In summary, we have derived the angular distribution of the final particles in the nonleptonic decay of the 
polarized charmed baryon  $\Lambda_c^+$, namely $\Lambda_c^+\to K^- \, \Delta (1232)^{++} \to K^- \, p \, \pi^+$. 
Several asymmetries have been proposed which are convenient for experimental determination of the 
components of the $\Lambda_c^+$ polarization vector. These asymmetries can be useful in the future measurements 
of magnetic and electric dipole moments of the charmed baryon $\Lambda_c^+$  at the SPS and the LHC using technique of 
channeling of charged particles in bent crystals~\cite{Baryshevsky:2016,Fomin:2017,Botella:2017,Bagli:2017,Fomin:2019,Mirarchi:2019}. 
This is part of the Physics Beyond Colliders project~\cite{Redaelli:2018,Scandale:2019} with a fixed-target setup at CERN. 

We show that the precession angles of the baryon polarization ${\cal \vec{P}}$ after passing of the baryon through 
the bent crystal are related to ratios of asymmetries. In these ratios the magnitude of polarization ${\cal P}$ 
and asymmetry parameter $\alpha$ do not enter, which is convenient for measurement of the precession angles and thereby 
the magnetic and electric dipole moments of $\Lambda_c^+$.    

Further, we have calculated the decay rate and asymmetry parameter 
for $\Lambda_c^+\to K^- \, \Delta (1232)^{++}$ in the pole model of Refs.~\citen{Xu:1992, Xu_Kamal:1992}. 
The parameters of the model have been updated using the present experimental 
information~\cite{PDG:2018}. Results of calculation are in reasonable agreement with available data.    
As an additional test of the pole model, the rate of the strange baryon decay $\Omega^- \to K^- \, \Lambda$ has been estimated
and compared with experiment.

%%%%%%%%%%%%%%%%%%%%%%%%%%%%%%%%%%%%%
%%%%%%%%%%%%%%%%%%%%%%%%%%%%%%%%%%%%%

\section*{Acknowledgments}
This work was partially conducted in the scope of the IDEATE International Associated Laboratory (LIA).
The authors acknowledge partial support by the National Academy of Sciences of Ukraine 
via the program ``Support for the development of priority areas of scientific research'' (6541230).
A.Yu.K. is grateful to Achille Stocchi for useful discussions.

%%%%%%%%%%%%%%%%%%%%%%%%%%%%%%%%%%%%%%%%%%%%%%%%%%%%%%%
% BIBLIOGRAPHY
%%%%%%%%%%%%%%%%%%%%%%%%%%%%%%%%%%%%%%%%%%%%%%%%%%%%%%%


\begin{thebibliography}{99}

\bibitem{Galanti:2015}
M.~Galanti, A.~Giammanco, Y.~Grossman,  Y.~Kats, E.~Stamou and J.~Zupan, 
% Heavy baryons as polarimeters at colliders,
J.\ High\ Energy\ Phys.\ {\bf 11}, 067 (2015).
%[arXiv:1505.02771v2 [hep-ph]]

\bibitem{Korchin:2016}
A.~Yu.~Korchin and V.~A.~Kovalchuk, 
% Decay of the Higgs boson to $\tau^-\tau^+$ and non-Hermiticity of the Yukawa interaction, 
Phys.\ Rev.\ D {\bf 94}, 076003 (2016). 
% DOI: 10.1103/PhysRevD.94.076003, arXiv:1607.02827v2 [hep-ph]

\bibitem{Jezabek:1992}
M.~Jezabek, K.~Rybicki, R.~Rylko, 
% Experimental study of spin effects in hadroproduction and decay of $\Lambda^+_c$,
Phys.\ Lett.\ B {\bf 286}, 175 (1992). 
%DOI: 10.1016/0370-2693(92)90177-6

\bibitem{Aitala:2000}
 E.~M.~Aitala {\it et al.} [E791 Collaboration],
% Multidimensional resonance analysis of $\Lambda_c^+ \to p \, K^- \pi^+$,
  Phys.\ Lett.\ B {\bf 471}, 449 (2000).
	%DOI: 10.1016/S0370-2693(99)01397-0 e-Print: hep-ex/9912003

\bibitem{Baryshevsky:2016} 
  V.~G.~Baryshevsky,
 % The possibility to measure the magnetic moments of short-lived particles (charm and beauty baryons) at LHC and FCC 
% energies using the phenomenon of spin rotation in crystals,
  Phys.\ Lett.\ B {\bf 757}, 426 (2016).

\bibitem{Fomin:2017}
A.~S.~Fomin, A.~Yu.~Korchin, A.~Stocchi, O.~A.~Bezshyyko, L.~Burmistrov, S.~P.~Fomin, I.~V.~Kirillin, 
L.~Massacrier, A.~Natochii, P.~Robbe {\it et al.}, 
% Feasibility of measuring the magnetic dipole moments of the charm baryons at the LHC using bent crystals, 
J.~High~Energy~Phys.\ {\bf 08}, 120 (2017).

\bibitem{Botella:2017}
  F.~J.~Botella, L.~M.~Garcia Martin, D.~Marangotto, F.~M.~Vidal, A.~Merli, 
	N.~Neri, A.~Oyanguren and J.~R.~Vidal,
  % On the search for the electric dipole moment of strange and charm baryons at LHC,
  Eur.\ Phys.\ J.\ C {\bf 77}, no. 3, 181 (2017).

\bibitem{Bagli:2017} 
  E.~Bagli, L.~Bandiera, G.~Cavoto, V.~Guidi, L.~Henry, D.~Marangotto, F.~Martinez Vidal, 
	A.~Mazzolari, A.~Merli, N.~Neri {\it et al.}, 
  % Electromagnetic dipole moments of charged baryons with bent crystals at the LHC,
  Eur.\ Phys.\ J.\ C {\bf 77}, no. 12, 828 (2017).

\bibitem{Fomin:2019} 
 A.~S. Fomin, S.~Barsuk, A.~Yu.~Korchin, V.~A.~Kovalchuk, E.~Kou, 
M.~Liul, A.~Natochii, E.~Niel, P.~Robbe and A.~Stocchi,
  % The prospect of charm quark magnetic moment determination,
  Eur.\ Phys.\ J.\ C {\bf 80}, no. 5, 358 (2020).
	%arXiv:1909.04654 [hep-ph].

\bibitem{Mirarchi:2019} 
  D.~Mirarchi, A.~S.~Fomin, S.~Redaelli and W.~Scandale,
  % Layouts for fixed-target experiments and dipole moment measurements of short-living baryons using bent crystals at the LHC,
  arXiv:1906.08551 [physics.acc-ph].

\bibitem{Bernotas:2013}
A.~Bernotas and V.~\v{S}imonis,  
% Magnetic moments of heavy baryons in the bag model reexamined,
Lith.\ J.\ Phys.\ {\bf 53}, 84 (2013). 

\bibitem{Simonis:2018} 
  V.~\v{S}imonis,
 % Improved predictions for magnetic moments and M1 decay widths of heavy hadrons,
  arXiv:1803.01809 [hep-ph].

\bibitem{PDG:2018}
 M.~Tanabashi {\it et al.} (Particle Data Group), 
% Review of Particle Physics, 
Phys. \ Rev. \ D {\bf 98}, 030001 (2018) and 2019 update.

\bibitem{Belle:2014} A.~Zupanc {\it et al.} [Belle Collaboration], 
% Measurement of the branching fraction $B(\Lambda^+_c\to p\,K^-\pi^+)$, 
Phys.\ Rev.\ Lett.\ {\bf 113}, 042002 (2014). 
% arXiv:1312.7826.

\bibitem{BESIII:2016} M.~Ablikim {\it et al.} [BESIII Collaboration], 
% Measurements of absolute hadronic branching fractions of the $\Lambda^+_c$ baryon,
Phys.\ Rev.\ Lett.\ {\bf 116}, 052001 (2016). 
% arXiv:1511.08380.

\bibitem{Xu:1992}
  Q.~P.~Xu and A.~N.~Kamal,
  % Cabibbo favored nonleptonic decays of charmed baryons,
  Phys.\ Rev.\ D {\bf 46}, 270 (1992).

\bibitem{Xu_Kamal:1992}
  Q.~P.~Xu and A.~N.~Kamal,
  % The nonleptonic charmed baryon decays: $B_c \to B (\tfrac{3}{2}^+, \ \mathrm{decuplet}) + P(0^-)$ 	or $V(1^-)$,
  Phys.\ Rev.\ D {\bf 46}, 3836 (1992).

\bibitem{Korner:1992}
  J.~G.~K\"{o}rner and M.~Kr\"{a}mer,
  % Exclusive nonleptonic charm baryon decays,
  Z.\ Phys.\ C {\bf 55}, 659 (1992).

\bibitem{Pilkuhn:1979}
H.~M.~Pilkuhn, {\it Relativistic particle physics} (Springer-Verlag, New York, 1979).

\bibitem{Koch:1980}
  R.~Koch and E.~Pietarinen,
  % Low-energy pi N partial wave analysis,
  Nucl.\ Phys.\ A {\bf 336}, 331 (1980).
	
\bibitem{Fomin:2018}
  A.~S.~Fomin, A.~Yu.~Korchin, A.~Stocchi, S.~Barsuk and P.~Robbe,
  % Feasibility of $\tau$-lepton electromagnetic dipole moments measurement using bent crystal at the LHC,
  J.\ High\ Energy\ Phys. {\bf 03}, 156 (2019).
 %[arXiv:1810.06699 [hep-ph]].

\bibitem{Bargmann:1959}  V.~Bargmann, L.~Michel and V.~L.~Telegdi,
  % Precession of the polarization of particles moving in a homogeneous electromagnetic field,
  Phys.\ Rev.\ Lett.\  {\bf 2}, 435 (1959).

 \bibitem{Berestetskii:1982}
V.~B.~Berestetskii, E.~M.~Lifshitz and L.~P.~Pitaevskii,    {\it Quantum electrodynamics} (Pergamon Press, Oxford U.K., 1982).

\bibitem{Jackson:1999} J.~D.~Jackson,  {\it Classical electrodynamics},  3rd edn. (John Wiley, 1999).

\bibitem{Lyuboshits:1980} 
  V.~L.~Lyuboshits,
  % The spin rotation at deflection of relativistic charged particle in electric field,
  Sov.\ J.\ Nucl.\ Phys.\  {\bf 31}, 509 (1980) \  [Yad.\ Fiz.\  {\bf 31}, 986 (1980)].

\bibitem{Kim:1983}
I.~J.~Kim, 
% Magnetic moment measurement of baryons with heavy flavored quarks by planar channeling through bent crystal,
Nucl.\ Phys.\ B {\bf 229}, 251 (1983).

\bibitem{Rosner:2010} 
  J.~L.~Rosner and S.~Stone,
 % Leptonic decays of charged pseudoscalar mesons,
  arXiv:1002.1655 [hep-ex].

\bibitem{Sokhoyan:2015} 
  V.~Sokhoyan {\it et al.} [CBELSA/TAPS Collaboration],
 % High-statistics study of the reaction $\gamma p \to p \; 2\pi^0$,
  Eur.\ Phys.\ J.\ A {\bf 51}, no. 8, 95 (2015) \  [Erratum:\ {\it ibid.} {\bf 51}, no. 12, 187 (2015)].

\bibitem{Konig:1993} 
  B.~K\"{o}nig, J.~G.~K\"{o}rner and M.~Kr\"{a}mer,
 % On the determination of the \ $b \to c$ \ handedness using nonleptonic $\Lambda_c$ decays,
  Phys.\ Rev.\ D {\bf 49}, 2363 (1994).

\bibitem{Stech:1990}
  B.~Stech and Q.~P.~Xu, 
% Hyperon decays and quark-quark correlations,     
	Z.\ Phys.\ C {\bf 49}, 491 (1991).

\bibitem{Redaelli:2018} 
  S.~Redaelli, M.~Ferro-Luzzi and C.~Hadjidakis,
 Studies for future fixed-target experiments at the LHC in the framework of the CERN physics beyond colliders study,
	in 9th International Particle Accelerator Conference,
  doi:10.18429/JACoW-IPAC2018-TUPAF045.

\bibitem{Scandale:2019} 
  W.~Scandale, F.~Cerutti, L.~S.~Esposito, M.~Garattini, S.~Gilardoni, S.~Montesano, 
	R.~Rossi, L.~Burmistrov, S.~Dubos, A.~Natochii {\it et al.},
  % Double-crystal setup measurements at the CERN SPS,
  arXiv:1909.02756 [physics.acc-ph].


\end{thebibliography}
\end{document}